\documentclass[10pt,a4paper]{article}
\usepackage{amsmath}
\usepackage{amsthm}
\usepackage{epsfig}
\usepackage{latexsym}
\usepackage{xcolor}
\usepackage{pdflscape}
\numberwithin{equation}{section}
\usepackage{graphicx}
\usepackage{dcolumn}

\title{\Large A tribute to Marian Smoluchowski's legacy on soft grains assembly and hydrogel formation\footnote{
 Presented in part at XXX Marian Smoluchowski Symposium, Cracow, September 4, 2017}%
}
\author{  Adam Gadomski$^1$,  Natalia Kruszewska$^1$, Piotr Be{\l }dowski$^1$\\ Bogdan Lent$^2$ \\Marcel Ausloos$^{3,4}$}

 \date{
$^1$ Group of Modeling of Physicochemical Processes,\\
Institute of Mathematics \& Physics, UTP University of Science \& Technology,  Kaliskiego 7, PL--85796 Bydgoszcz,  Poland\\ \vskip0.5cm
$^2$ Faculty of Management, UTP University of Science \& Technology,  Kaliskiego 7, PL--85796 Bydgoszcz,  Poland\\
\vskip0.5cm
$^3$ GRAPES\footnote{Group of Researchers for Applications of Physics in Economy and Sociology}$\;$,
rue de la Belle Jardini\`ere 483, \\B-4031, Angleur, Belgium\\ 
$^4$ School of Business, University of Leicester, University Road, Leicester  LE1 7RH, United Kingdom \\}
\begin{document}
 \maketitle


\begin{abstract}
The paper compares the statistical description of physical-metallurgical processes and ceramic-polycrystalline evolutions, ter\-med the normal grain growth (NGG), as adopted to soft- and chemically-reactive grains, with a Smoluchowski's population-constant kernel cluster-cluster aggregation (CCA) model, concerning irreversible chemical reaction kinetics. The former aiming at comprehending, in a semi-quan\-ti\-ta\-tive way, the volume-conservative (pressure-drifted) grain-growth process which we propose to adopt for hydrogel systems at quite low temperature (near a gel point). It has been noticed, that by identifying the mean cluster size $<k>$ from the Smoluchowski CCA description with the mean cluster radius' size $R_D$,  from the NGG approach of proximate grains, one is able to embark on equivalence of both frameworks, but only under certain conditions. For great enough, close-packed clusters, the equivalence can be obtained by rearranging the time domain with rescaled time variable, where the scaling function originates from the dispersive (long-tail, or fractal) kinetics, with a single exponent equal to $d+1$ (in $d$-dimensional (Euclidean) space).  
This can be of interest for experimenters, working in the field of thermoresponsive gels formation, where crystalline structural predispositions overwhelm.
\end{abstract} 
  
 \maketitle
 


\section{Introduction}\label{intro}

In 1916 Marian Smoluchowski proposed a case of (populationally fixed) constant-kernel cluster-cluster aggregation (CCA), for which it is manageable to find analytically, by employing scaling arguments, a solution in terms of the cluster size ($k$) distribution function, $n(k)$ \cite{smol,jullien}. By applying this scaling function it is then possible to get, within the long time limit, the results for the mean cluster size $<k>$ and the total number of the clusters $N_c$, both scalable in terms of time with a single exponent, denoted by $\gamma$ \cite{jullien}. The clustering arguments, first introduced by Smoluchowski \cite{smol}, are easily applicable to a statistical description of physical-metallurgical processes and ceramic-polycrystalline evolutions, termed the normal grain growth (NGG), in which bigger clusters grow at the expense of their smaller neighboring counterparts due to preferentially capillary conditions \cite{mulh}. The NGG, and their dynamics, can be expressed in $d$-dimensional (Euclidean) space. 
In this study, it is proposed, that upon identifying $<k>$ from the Smoluchowski CCA description with the mean cluster radius' size $R_D$,  from the NGG approach of proximate soft-and-reactive grains, one is able to embark on their equivalence. However, a few assumptions are necessary. The most important is appearing fully feasible when rearranging the time domain by substituting $t$ in a way such that a new rescaled time variable $\tau(t)$ is given by a definite integral in $[0,t]$ upon $d\tau(t)=dt/f(t)$, with an adjustable (albeit auxiliary) function $f$, coming out from the dispersive or long-tail, or fractal kinetics' arguments, which are endemic in condensed media \cite{plonka}. The arguments may, at least qualitatively, concern biomembranes dynamics. They can also contribute to nucleation-growth processes in soft-matter conditions \cite{AG1,agchp} as well as to hydogels with prevailing microcrystalline inclusions \cite{Ren}.

Hydrogels are example of microgels defined as viscoelastic systems classified to be certain  intermediates between polymer chains, viz coils, and the so-called macrogels, such as gelatine or yoghurt \cite{tp,natalia}. They are often chemically prepared to be designated as two-component systems. They consist of mixed solute and solvent phases in which solvent molecules interact with solute particles composed of polymer chains, and their aggregates, prone to behave in a network-like manner, but with a prevailing number of microcrystalline domains included inside their structure \cite{Ren,tp}.
By virtue of their complicated intimate interaction map they suffer difficulties in view of reaching a thermodynamic equilibrium. Their viscoelastic properties undergo some structural-geometric changes in the course of temperature and time. Those changes, such as microgel volume's expansion, emerge as a result of their sensitivity to temperature conditions. The conditions, in turn, are able to alter, in the course of time, the quality of the solvent making it either good or poor in terms of its affinity to solute molecules \cite{atak1}. The good solvent conditions cause the polymer chains to expand in space due to solvent molecules absorption, technically called occlusion. If the solvent molecules, met mostly in either inappropriate (harsh) or typically low temperature conditions, act as badly as possible while interacting with a polymer chain, making it shrunken or obstructed in its capability of gaining more space around, then an opposite physicochemical scenario prevails. The former is termed a coil effect whereas the latter is known as a globule counter-effect. The temperature, being a control parameter, establishes then a passage between coil and globule by decisively altering the solvent conditions, a physicochemical scenario so well-described by the Flory-Stockmayer theory, and well envisaged by the  pivotal role played by the Flory-Huggins solute-solvent interaction energy parameter \cite{dimar,dimars}.

Therefore, the NGG theory of close-packed entropic systems can describe hydrogel formation only in a low temperature regime (and close to the gel point), where hydrogel grains are soft, connected with one another by means of weak bonds. After some critical temperature the structure starts to be loosely-packed and the system can not be described in terms of mean-field approach. 

The article is organized in the following way. By employing a cluster-cluster analogy of colloid type \cite{agchp,agjs}, in Sec. \ref{sol} we try to unravel a model sol-like (typically, non-isothermal) system, apparently under low temperature circumstances, thus, grasped in a low energy well. Such a system is virtually able to conserve its total volume (or, sometimes, area \cite{atak1}), and may remain nearly inactive as far as its overall spatial expansion is concerned. In Sec. \ref{timescal}, 
a Smoluchowski's populationally (up to $i+j=k$-value as compared with constant-volume modified NGG approach) constant-kernel CCA approach has been presented and compared with close-packed NGG theory. All assumptions has been introduced which are necessary to state the assertion about their equivalence. 
Sec. \ref{summ} provides conclusions. It also gives us an outlook of the  approach applied, emphasizing the fact that the analogy addressed suits truly well, at least in a qualitative manner, the nonergodic viscoelastic framework staying behind. It applies in particular to the ones of bioreactive gels and/or living-matter involving contexts where microgels with swollen microcrystalline domains exist.

\section{Sol-like model system at a low thermal energy well}\label{sol}
The NGG model is a  simple theoretical construct \cite{agjs,agchp}. It assumes that the role of clusters is played by hydrogel's polymers that absorbed a suitable fraction of the solvent molecules.
The polymer chains of hydrodynamic radius $R_h$, occluded by the molecules, constitute solvent-involving domains of the effective domain-occupation volume $v$ such that at a time $t$ both quantities are $t$--dependent. At a given temperature $T$ and in $d=3$ dimensional (Euclidean) space, they are obedient to a simple geometric proportionality relation
\begin{eqnarray} \label{simplegeom}
{v(t)} \sim {{R_{h}}^3(t)} ,
\end{eqnarray}
that tacitly postpones the form factor of, e.g. the hydrated polymer domain, provided that we confine ourselves to hydrogels \cite{tp}.\\
Let us assume, that we have to do with a semi-concentrated polymer solution in which the solute and solvent coexist at a relatively low $T$ such that the solvent, viz water, is unable to cause the polymer globules to become coils. It is because it is not capable of penetrating the polymer's interior in order to swell the chain or to help the polymer expand into the neighboring territory. But under such circumstances the poorly swollen polymers are able to (a) diffuse under dynamic structural confinement, both, in terms of their mass-center motions and rotational movements; (b) interact with each other yielding dimers, oligomers, and some aggregates, finally. Due to low $T$ and quite high concentration conditions, their motions are fairly restricted. It can be foreseen that they will form then a more or less cellular microstructure with well separated but poorly hydrated polymer domains. The microstructure would to a first approximation be reminiscent of  a sol phase since the domains are rather immobilized and less reactive, according to their reactive encounters that are anticipated to be too small. The exchange of matter between neighboring domains occurs due to local pressure differences, sometimes accompanied by the corresponding structural rearrangements of diffusive nature \cite{agchp}.\\
Such domains resemble tightly built clusters or even 'soft' grains/assemblies that might have appreciably well defined surface-tension factor. This information can be addressed to the system (stochastic) dynamics in terms of the current $J(v,t)$ along the "reaction coordinate" ($v$) which, after adopting its form from a Smoluchowski-type model of CCA and its isothermal evolution \cite{agchp,agjs}, can be proposed as
\begin{eqnarray} \label{current}
J(v,t) =  -{{\partial\over{\partial v}}{\Big( D(v)\phi(v,t)\Big)}} ,
\end{eqnarray}
where $\phi(v,t)$ is the probability density of finding a domain of volume $v$ at time $t$. The quantity $dn=\phi(v,t)dv$ represents the relative number of polymer domains or "grains"  having the volumes kept in the interval $\left[v,v+dv\right]$. The $v$-dependent (or, state-dependent) diffusion function, $D(v)$, indicates quantitatively a colloid type cluster formation \cite{agjs}, thus, for Euclidean space dimension $d=3$ it is provided by
\begin{eqnarray} \label{diffusionf}
{D(v)} =  {D_ov^{2/3}} ,
\end{eqnarray}
where $D_o$ - a proper dimension keeping constant. Note that, because of Eq. (\ref{simplegeom}), $v^{2/3}\propto {R_{h}}^2$, which means, that $D$ from Eq. (\ref{diffusionf}) is designed as being proportional to the domain surface, $s_D$, i.e. $s_D \propto {R_{h}}^2$ applies.\\
To reveal what a content of physical message is included in Eq. (\ref{current}), let us proceed with the differentiation over $v$ at the right-hand side of the Eq. (\ref{current}). After so performing, it is expressed by 
\begin{eqnarray} \label{current2}
J(v,t) =  -{{D'(v) \phi(v,t)} - D(v){\partial\over{\partial v}}{\phi(v,t)}} ,
\end{eqnarray}
in which $D'(v)=(2D_o/3)v^{-1/3}$. Notice, however, that $2v^{-1/3} \sim 2/{R_h}$, and the prefactor $D_o/3$ keeps again track of proper physical units. The quantity $\kappa_D=2/{R_h}$ stands for twice the mean curvature of the sol particle viz the shrunken polymer domain of globular propensity. Each one of such domains conforms to some pressure difference $\Delta\pi_D$ between the external and internal parts of the domain, following the Kelvin-Laplace law, namely
\begin{eqnarray} \label{K-L}
{\Delta\pi_D} =  {\sigma\kappa_D} ,
\end{eqnarray}
providing the surface tension of the domain circumference obeyed: $\sigma\propto D_o$. Of course $\sigma=\sigma(T)$ ought to be taken for granted. The pressure $\Delta\pi_D$, being comparably to $v$, a stochastic variable, changes over time within the sol-like but semi-concentrated (i.e. well packed or relatively dense \cite{agchp}) matrix during its evolution. Moreover, and still within our approximate reasoning offered (see, Eq. (\ref{simplegeom}) and discussion below Eq. (\ref{current2})), the instantaneous pressure is naturally involved via the Kelvin-Laplace law of micro-capillarity\footnote{The linear size of  polymer-solvent domains of volume $v$ should remain comparable with $100 {nm}$ \cite{tp}, thus, belonging to the submicron scale.} in the current $J(v,t)$
\begin{eqnarray} \label{currentPi}
J(v,t) =  -{{{ \Delta\pi_D} \phi(v,t)} - D(v){\partial\over{\partial v}}{\phi(v,t)}} ,
\end{eqnarray}
when one accepted that $\sigma= {D_o/3}$ applies. An explanation of it can also be found elsewhere \cite{agcrt}.\\
The structural current is then taken to obey the continuity equation which is the evolution equation for the probability density $\phi(v,t)$. The continuity equation reads
${{{\partial\over{\partial t}}\phi(v,t)} + {\partial\over{\partial v}}{J(v,t)}} = 0$.\\
It should be completed by the so-called normal boundary conditions of absorbing type $\phi(v=0,t)=\phi(v=\infty,t)=0$, meaning that no grains' magnitude prevails during the system's evolution \cite{agjs}. The initial condition has to be selected  as well.  The simplest  selection can be a delta Dirac distibution \cite{agchp,agcrt}.\\
The approach offers three important measures of the pres\-sure-drif\-ted diffusion dynamics of the well packed sol-like and weakly reactive system. Two of them, $n(t)$-the average number of the domains, as well as $V(t)$, designating the total system volume, can be read out from
\begin{eqnarray} \label{moments}
m_i (t)={{{\int_0}^\infty}v^i \phi(v,t)dv},
\end{eqnarray}
where $n (t)=m_0 (t)$ and $V(t)=m_1 (t)$; $i=1,2,3,...$. The solution $\phi (v,t)$, following the variable-separation method,  has already been provided elsewhere \cite{agjs}. Third quantity of greatest concern is the average radius of the dehydrated globular domain, $R\equiv R_D (t)$, which is to be estimated based on a global vs. local volume geometrical relation
\begin{eqnarray} \label{radius}
 {V(t)}\simeq {n(t) {{R_D}^3 (t)}}.
\end{eqnarray}
Bear in mind that ${R_D}^3 (t)$ corresponds to the average volume of a single domain, and the averaging is performed as an  integration over $v$, where $v\in [0,\infty]$, cf. Eq. (\ref{moments}). Averaging $<{R_D}^3 (t)>$ in the spirit of Eq.(\ref{moments}) is supposed to be performed in a nearly fluctuationless regime (about mean-field approach), implying that $<{R_D}^3(t)>=<{R_D}(t)>^3$ \cite{agchp,PLA95}. It is, however, very consistent with CCA Smoluchowski approach presented in the Sec. \ref{timescal} (cf. Eq.(\ref{smol_eq})), provided that $\sum \longrightarrow \int$.
 
It is interesting to notice that the three key dynamic quantities do obey scaling laws at their asymptotic regimes for which $t\gg t_o$ ($t_o$, an initial instant).
First, the number of domains, $n(t)$, conforms to a scaling law of
\begin{eqnarray} \label{numberS}
 {n(t)}\sim {t^{-3/4}}.
\end{eqnarray}
Second, the volume $V(t)$ is expected  to obey a  constancy condition \cite{last}, thus it quasi-scales trivially with $t^o$, i.e.
\begin{eqnarray} \label{volumeC}
 {V(t)}={V(t_o)\to const.}
\end{eqnarray}
Third, the average radius, $R_D (t)$ scales as
\begin{eqnarray} \label{radiusS}
 {R_D (t)}\sim {t^{1/4}}.
\end{eqnarray}
 Note that in a $d$--dimensional space the scaling goes as ${R_D (t)}\sim {t^{1/{(d+1)}}}$. Realize that Eq. (\ref{radius}) is consistent with the scaling laws provided, cf. \cite{agchp,agjs}. Thus, according to Eq. (\ref{volumeC}) the sol system is envisaged as a conservative and non-expanding  because its total volume is a conserved quantity upon such low-energy thermal conditions or because of being trapped in a low-energy well. It resembles some stagnant and weakly reactive cellular network in which, however, the network eyes (domains) are nearly disjoint objects due to some non-negligible $\Delta\pi_D$--s distributed uniformly over the system under study the values of which become constant.
It is profitable to look at the distribution of $\Delta\pi_D$--s over the total volume $V$. Since $\Delta\pi_D\equiv \Delta\pi_D (t)$ can be taken from Kelvin-Laplace law, Eq. (\ref{K-L}), and because $\kappa_D = 2v^{-1/3}$, one is able to provide an equivalent of Eq. (\ref{K-L}) to be rewritten as $\Delta\pi_D (v) = 2\sigma v^{-1/3}$. From it one infers that $v = {(2\sigma /\Delta\pi_D (v))}^3$. One might then evaluate, based on Eq. (\ref{moments}) and Eq. (\ref{volumeC}), a ensemble-averaged specific quantity ${\langle 2\sigma/\Delta\pi_D (v)\rangle}^3 = m_1 (t)=V(t)=V(t_o)$. To be precise, the average, owing to the statistical uniformity of the system, reads ${\langle 2\sigma/\Delta\pi_D (v)\rangle}^3 \equiv {(2\sigma)}^3{\langle\Delta\pi_D (v)\rangle}^{-3}=V(t_o)=const.$ Because the domain's surface tension $\sigma$ is assumed to be independent of $t$ during the evolution (the domain shells are characterized by mainly $T$--dependent surface tensions), one is able to address in full the constancy of  $\langle\Delta\pi_D (v)\rangle$ by providing the following
\begin{eqnarray} \label{pressureC}
 {\langle \Delta\pi_D (v)\rangle}={2\sigma\over {{[V(t_o)]}^{1/3}}}={const.}
\end{eqnarray}
Thus, the overall $\langle\Delta\pi_D (v)\rangle$ takes on a well appreciated constant value.
One may also anticipate that an internal mechanical stress assigned to the polymeric system at the late-stage limit \cite{vK} will distribute uniformly in very similar way too $\Delta\pi_D$--s do. Thus, for an ideal (equilibrium) cellular network in a $2D$ space, envisaged by a honeycomb microstructure, the mechanical stress would distribute over the triple junctions crossing points nearly at the angle of $2\pi/3$. In certain  bubbles-containing (or, soap froths) quite analogous systems, however, the very circumstance  could be different \cite{blast}.

\section{Argumentation for rescaling the time variable}\label{timescal}

\begin{figure}[htb]
\begin{center}
\includegraphics[width=0.9\textwidth]{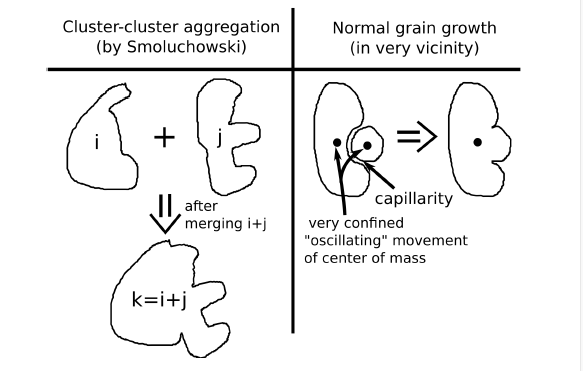}
\caption{Main concepts coming out from both: Smoluchowski's CCA and soft-and-reactive NGG, frameworks. In the former, aggregates consisted of $i$ units merging with aggregates of $j$ units and this way $k$-units aggregates are formed \cite{smol,jullien}. In NGG (grains in very close vicinity), characteristic of metallurgical grain growth, smaller grains are captured by greater ones - this way grains become bigger and their centers of mass are shifted \cite{agchp,PLA95}. }\label{sketch}
\end{center}
\end{figure}

After presenting the standard approach to grain- or soft domain-growth of aggregates (NGG), 
let us here - provided that both approaches have much in common -  compare it to Smoluchowski’s CCA framework with a kernel which is dependent only on time \cite{smol}. 
A graphical sketch of a main concept, standing behind stating the equality of both frameworks (CCA \& NGG), is presented in the Fig. \ref{sketch}.

Smoluchowski, in his CCA approach, claimed that for irreversible chemical reaction, $[i]+[j]\longrightarrow[i+j]$, with kinetic constant (kernel) $K_{ij}$ of the reaction, a rescaled-time $\tau$ evolution equation of concentration, $c_k$, of $k$ aggregates ($k=i+j$), can be given by
\begin{eqnarray} \label{smol_eq}
 \frac{dc_k}{d\tau}=\frac{1}{2}\sum_{i+j=k}K_{ij}c_{i}c_{j}-\sum_{i}K_{ik}c_ic_k,
\end{eqnarray}
where $c_i$ and $c_j$ are concentrations of the ingredients \cite{smol,jullien}. One-half on the r.h.s. of Eq. (\ref{smol_eq}) means that binary (merging) interactions cannot be counted twice.
The equation can be written in terms of the number of clusters of $k$ particles $n_k=c_kV$, where $V$ is the total volume of the solution. Smoluchowski proposed a solution of the equation, in case of a populationally constant kernel ($K_{ij}=\kappa=const(i,j).$). It takes a scaling form
\begin{eqnarray} \label{smol_eq_sol}
 n_k= N\frac{(\kappa \frac{N}{2}\tau )^{k-1}}{(1+\kappa \frac{N}{2}\tau)^{k+1}},
\end{eqnarray}
where $N$ is the total number of particles (see, Sec. 6.2 in \cite{jullien} for details on the method of solving the Eq. (\ref{smol_eq})). 
The total number of clusters reads
\begin{eqnarray} \label{smol_totNc}
 N_c= \frac{N}{1+\frac{N}{2}\kappa \tau}.
\end{eqnarray}
Notice, that constancy of $\kappa$ means independence of $i$ and $j$ but not of time $\kappa=\kappa(\tau)=>\kappa=\kappa_cf(\tau)$, where $\kappa_c$ is a real constant and $f$ is an aging viz prolonged reactivity expressing (dispersion) function.
After assuming a large time regime, the size distribution function from Eq. (\ref{smol_eq_sol}) can be written in a following scaling form with a Boltzmann type broken valued argument in an adjustable function F:
\begin{eqnarray} \label{smol_scal_nk}
 n_k= k^{-\mu}F(\frac{k}{\tau^\gamma}),
\end{eqnarray}
with $\mu=2$ and $\gamma=1$.
From the formalism presented above, one can obtain time scaling rules on mean cluster size $<k>$ and number of clusters $N_c$ \cite{jullien}:
\begin{eqnarray} \label{k_scal}
 <k(\tau)>\propto \tau^\gamma
\end{eqnarray}
and
\begin{eqnarray} \label{Nc_scal}
 N_c(\tau)\propto \tau^{-\gamma}.
\end{eqnarray}

Our statement of equivalence of the Smoluchowski's CCA description with the standard growth of grains rely upon identifying $<k>$ from the Smoluchowski's description with the mean cluster radius' size $R_D$ (see, Eq.(\ref{radiusS}) in Sec. \ref{sol}), and by taking the condition of $k >> 0$. Notice, that the sum from Eq. (\ref{smol_eq}), due to the condition of $k>>0$, can be replaced by the corresponding integral, which lies within the spirit of the comparison between the two approaches, cf.  Eq.(\ref{moments}) accompanied by Eq. (\ref{radius}) in which forms the integral expressions are involved.
The crucial assumption, however, that assures the equivalence claimed ($<k>\longrightarrow R_D$), appears to be fully feasible when rearranging the time domain \cite{PLA95} by substituting $t$ in such a way: 
\begin{eqnarray} \label{t_scal}
 \tau=\tau(t)=\int\limits_0^t f^{-1}(t')dt',
\end{eqnarray}
with an adjustable function $f$, coming out from the dispersive (fractal- long-tail) kinetics arguments \cite{plonka}. 
This aging function, $f$, should be modelled in a scalable form as: $f(\tau)=\frac{const.}{\gamma_{GG}}\tau^{\gamma_{GG}-1}$, where $\gamma_{GG}=d+1$.
The scaling exponent, presented in the statistical moments, $\gamma=1/(d+1)$, since the asymptotic scaling rule for $N_c$ goes via a simple logarithmic depiction as: $\ln{N_c}\sim-\gamma\ln{\tau}$ (cf. Eqs (\ref{radiusS}), (\ref{k_scal}), (\ref{Nc_scal})). 

The kernel function $\kappa$ can be time dependent because of some additional energy provided, in a controlled way, in the system by very carefully increasing $T$ gradually in time, say, from some $T$ to a $T+\langle\Delta T\rangle$, wherein $\langle\Delta T\rangle >0$ very slightly is an averaged temperature step associated with the temperature's increase \cite{ph}.
In NGG approach, ${\langle\Delta\pi_D\rangle}$ depends on temperature because it is proportional to surface tension $\sigma$ which is fairly dependent on temperature. Thus, if $T$ is a function of time, then ${\langle\Delta\pi_D (t)\rangle}$ also. Both approaches (CCA and NGG), however, lost their "compatibility" at some critical point $T_c$ where close-packed regime in grain growth is relieved by loosely-packed one, what is characteristic of sol-gel phase change. It is due to the fluctuations of $R_D$ (cf. explanation under Eq. (\ref{radiusS})). Thus, soft-and-reactive NGG fairly close-packed description can most likely  be adopted for certain hydrogel systems \cite{Ren}.

\section{Conclusions}\label{summ}

In this study, a kinetic-thermodynamic depiction of a model hydrogel formation, with soft-and-reactive crystalline inclusions, has been unveiled in terms of a statistical-thermodynamical concept \cite{agchp,agjs,agcrt,PLA95}. To achieve this goal, a Smoluchowski-type CCA approach to the drifting and diffusive nature of the system has been adopted for modeling semi-quantitatively 
an expansion of proximate grains in time. The modeling has been performed with the aim of uncovering some basic trends of the hydrogel formation, which can be found also in \cite{tp,Ren}.

The conditions of dispersive kinetics are to be seen as indispensable upon identifying the basic domain-growth soft material systems (such as those of Langmuir-Blodgett type) with the classic Smoluchowski's CCA with time dependent kernel, being in the same time independent of the number of molecular constituents of the clusters upon clusters' absorption-desorption conditions applied. The equality of NGG and CCA approaches, can be stated with following assumptions: (i) the mean cluster size $<k>$ from CCA (within the long time limit) is identified with the average radius $R_D$ from NGG, which appears to be true when rearranging the time domain with rescaled time variable $\tau(t)$, given by a definite integral in $[0,t]$ upon $d\tau(t)=dt/f(t)$, with an adjustable function $f$, coming out from the dispersive kinetics' arguments \cite{plonka};
(ii) $k$ should be great enough to allow the sum from Eq. (\ref{smol_eq}) to be replaced by the corresponding integral, which lies within the spirit of the comparison between the two approaches, cf.  Eq.(\ref{moments}) accompanied by Eq. (\ref{radius}) in which forms the integral expressions are involved;
(iii) the time scaling exponent, presented in the statistical moments in both frameworks, $\gamma=1/(d+1)$ - it is only true for proximate grains (close-packed structures) where fluctuations of $R_D$ are close to zero (about mean-field approach).
The exponent $\gamma=1/4$ (for $d=3$ dimensional space) involved in the scaling relation (Eq. (\ref{radiusS})) keeps the signature of $(d+1)$-involving random close packing, a measure characteristic of a $d$--dimensional geometrical-physical space upon confinement \cite{AGRubi}. 

A certain novelty coming out from our statistical moments involving approach, see Eq. (\ref{moments}), appears to be a quite precise estimation of the average Laplace's pressure, $\langle\Delta\pi_D\rangle$, which turns out to be a constant value (Eq. (\ref{pressureC})) for the volume-conservative sol-like phase upon approaching gelation critical point \cite{tp}. However, it becomes $t$-dependent when one provides additional energy into system causing increasing of the temperature $T(t)$. 

Such a volume-conservative description of grain growth can be applicable to hydogels with prevailing microcrystalline inclusions \cite{Ren} also with very sensitive pH vs. temperature (implicitly, time-temperature sensitive; cf. Fig. 3b in \cite{Wang}) nanocomposite's expressions. Also, thermoresponsive gels with overwhelming crystalline structural predispositions (such as in \cite{Taylor}), commencing from the molecular level first, have to be invoked as a working example here. However, usage of the description is limited to the low-temperature regime, close to gel point.

To summarize, the main and very novel finding of this study is to convince the reader on reconciling that the $k$-fixed "constant" kernel celebrated approach by Marian Smoluchowski \cite{smol,jullien} can be recast from the dispersive viz soft-and-reactive NGG (but processing time rescaled) material formation \cite{plonka,AG1,agchp,PLA95}, provided that both approaches work within the realm of almost volume’s fluctuationless (stationary) regimen. 

It is also worth mentioning that at least in two areas of the approach employed, a dynamic and network-involving scenarios of microgel type (with swollen microcrystalline domains)  emerge inevitably. First, in the biophysical area of ultralow friction and facilitated lubrication of articular cartilage(s), see \cite{AC}, wherein the hyaluronic-acid, network-like constructs respond synergistically to the external load's action. Second, when within a cell the (anomalously bioreactive) metabolic pathway spreads out over its complex viscoelastic interior in intimately networking, and fairly dynamically organized manners, see \cite{TN}.

The presented model - upon identifying that the crystalline material insertions grow or rather swell uniformly - can qualitatively mimic malignancies and their tumor-like growth in a virtually  active matter, provided that we are able to override the fixed-population and volume constancy limits characteristic of both (comparative) approaches under study. In \cite{AG2017} it has been suggested that one has to embark on a fluctuational and viscoelastic clustering effect encountered in an 'active brainy' viz nearly constant-volume matter. For example, it is believed to change the emotions of an individual, and alter decision-making conditions, provided that structure-property and functional material unification is occurring. However, any discussion of time domain, especially the one due to mental hesitation involvement, becomes elusive and prone to interpretation, cf. \cite{AGLent}.
Especially, the decision making in leader-type biased personality is comprehended as some cognition complex task,
to be meaningfully simplified in physical, i.e. dispersive clustering involving, neurophysical terms \cite{AG2017,AGLent}.

\section*{Acknowledgment} A support of the present study by BS 39/2014 (UTP Bydgoszcz) is to be emphasized. AG benefited much from preliminary discussions with Prof. T. Wysocki (Nebraska Lincoln). Technical assistance of Mrs. H. Przewo{\'z}niak (UTP) is acknowledged.


\end{document}